\documentclass{aa}  

\usepackage{graphicx}
\usepackage{txfonts}
\usepackage{rotating}
\usepackage{adjustbox}
\usepackage{placeins}

\newcommand{\mgii}{\ion{Mg}{II}}
\newcommand{\civ}{\ion{C}{IV}}
\newcommand{\feii}{\ion{Fe}{II}}
\newcommand{\cii}{[C\,II]}

\usepackage{hyperref}
\begin{document}

   \title{XQR-30: Black Hole Masses and Accretion Rates of 42 $z\gtrsim 6$ Quasars} 
   \titlerunning{XQR-30: Black Hole Masses and Accretion Rates}
   \authorrunning{Mazzucchelli et al.}

   \author{C. Mazzucchelli\inst{1}\thanks{email: chiara.mazzucchelli@mail.udp.cl},
          %\and
          %all
          M. Bischetti\inst{2}
          V. D'Odorico\inst{2,3,4},
          C. Feruglio\inst{2},
          J-T. Schindler\inst{5},
          M. Onoue\inst{6,7},
          E. Ba\~nados\inst{8}, %orcid: [0000-0002-2931-7824]
          G. D. Becker\inst{9}, %orcid: [0000-0003-2344-263X]
          F. Bian\inst{10}, %orcid: []
          S. Carniani\inst{3},  %orcid: [0000-0002-6719-380X]
          R. Decarli\inst{11}, %orcid: [0000-0002-2662-8803]
          A.-C. Eilers\inst{12}\thanks{Pappalardo Fellow}, %orcid: [0000-0003-2895-6218] 
          E. P. Farina\inst{13}, %orcid: [0000-0002-6822-2254]
          S. Gallerani\inst{3}, %orcid: []
          S. Lai\inst{14}, %orcid: [0000-0001-9372-461]  
          R. A. Meyer\inst{8}, %orcid: [0000-0001-5492-4522]
          S. Rojas-Ruiz\inst{8}, %orcid: [0000-0003-2349-9310]
          S. Satyavolu\inst{15}, %orcid: []
          B. P. Venemans\inst{5}, %orcid: [0000-0001-9024-8322]
          F. Wang\inst{16}, %orcid: []
          J. Yang\inst{16}, %orcid: []
          Y. Zhu\inst{9} %orcid: [0000-0003-3307-7525]
          %C. Ptolemy\inst{2}\fnmsep\thanks{Just to show the usage
          %of the elements in the author field}
          }
          
    \institute{
    $^{1}$Instituto de Estudios Astrof\'{\i}sicos, Facultad de Ingenier\'{\i}a y Ciencias, Universidad Diego Portales, Avenida Ejercito Libertador 441, Santiago, Chile.\\ %[C\'odigo Postal 8370191]
    $^{2}$INAF--Osservatorio Astronomico di Trieste, Via G.B. Tiepolo, 11, I-34143 Trieste, Italy\\
    $^{3}$Scuola Normale Superiore, P.zza dei Cavalieri 7, I-56126 Pisa, Italy\\
    $^{4}$IFPU--Institute for Fundamental Physics of the Universe, via Beirut 2, I-34151 Trieste, Italy\\
    $^{5}$Leiden Observatory, Leiden University, Niels Bohrweg 2, 2333 CA Leiden, Netherlands\\
    $^{6}$Kavli Institute for Astronomy and Astrophysics, Peking University, Beijing 100871, People’s Republic of China\\
    $^{7}$Kavli Institute for the Physics and Mathematics of the Universe (Kavli IPMU, WPI), The University of Tokyo, Chiba 277-8583, Japan\\
    $^{8}$Max Planck Institut f\"ur Astronomie, K\"onigstuhl 17, D-69117, Heidelberg, Germany\\
    $^{9}$Department of Physics \& Astronomy, University of California, Riverside, CA 92521, USA\\
    $^{10}$ESO, Vitacura Alonso de C\'ordova 3107, Vitacura, Casilla 19001, Santiago de Chile, Chile\\
    $^{11}$INAF - Osservatorio di Astrofisica e Scienza dello Spazio di Bologna, Via Gobetti 93/3, I-40129 Bologna, Italy\\
    $^{12}$MIT Kavli Institute for Astrophysics and Space Research, 77 Massachusetts Avenue, Cambridge, 02139, Massachusetts, USA\\
    $^{13}$Gemini Observatory, NSF’s NOIRLab, 670 N A’ohoku Place, Hilo, Hawai'i 96720, USA\\
    $^{14}$Research School of Astronomy and Astrophysics, Australian National University, Canberra, ACT 2611, Australia\\
    $^{15}$Tata Institute of Fundamental Research, Homi Bhabha Road, Mumbai 400005, India\\
    $^{16}$Steward Observatory, University of Arizona, 933 N Cherry Avenue, Tucson, AZ 85721, USA   
    }

   \date{Received September 15, 1996; accepted March 16, 1997}

  \abstract
   {We present bolometric luminosities, black hole masses and Eddington ratios for 42 luminous quasars at $z\gtrsim6$ using high signal-to-noise ratio VLT/X-Shooter spectra, acquired in the enlarged ESO Large Programme {\it XQR-30}. In particular, we derive bolometric luminosities from the rest-frame 3000 \AA\, luminosities using a bolometric correction from the literature, and the black hole masses by modelling the spectral regions around the \civ\,1549\AA\, and the \mgii\,2798\AA\, emission lines, with scaling relations calibrated in the local universe. We find that the black hole masses derived from both emission lines are in the same range, and the scatter of the measurements agrees with expectations from the scaling relations. The \mgii-derived masses are between $\sim$(0.8$-$12) $\times 10^{9} M_{\odot}$, and the derived Eddington ratios are within $\sim$0.13$-$1.73, with a mean (median) of 0.84 (0.72). By comparing the total sample of quasars at $z>5.8$, from this work and from the literature, to a bolometric luminosity distribution-matched sample at $z\sim1.5$, we find that quasars at high redshift host slightly less massive black holes which accrete slightly more rapidly than at lower$-z$, with a difference in the mean Eddington ratios of the two samples of $\sim$0.27, in agreement with recent literature work.}

   \keywords{(galaxies:) quasars: supermassive black holes --
              (galaxies:) quasars: emission lines --
              galaxies: high-redshift
               }

   \maketitle
%
%-------------------------------------------------------------------

\section{Introduction}

   Quasars are the most luminous, non transient sources in the universe, hence they can be observed at very early cosmic times, into the Epoch of Reionization at $z\gtrsim6$ (within the first billion years of the universe; e.g.\,\citealt{jiang2015}, \citeyear{jiang2016}, \citealt{banados2016}, \citealt{reed2019}, \citealt{matsuoka2019}), up to $z\sim$7.5 (e.g.\,\citealt{banados2018a}, \citealt{yang2020}, \citealt{wang-feige2021}; see \citealt{fan2022} for a recent review). They are already powered by supermassive black holes (SMBHs) in their centers ($M_{\mathrm{BH}}>10^{8} M_{\odot}$ e.g.\,\citealt{jiang2007}, \citealt{shen2019}), challenging models of early black holes formation and growth (e.g., \citealt{inayoshi2020} and \citealt{volonteri2021} for recent reviews), and already present evolved broad-line-regions (BLRs) with super-Solar metallicities (e.g.\,\citealt{kurk2007}, \citealt{lai2022}).
   In order to grow a billion-solar masses SMBH by $z\gtrsim6$, models require either a ``light'' ($\sim10^{2} M_{\odot}$) SMBH seed undergoing rapid super-Eddington accretion episodes, or a ``heavy'' ($\sim10^{5} M_{\odot}$) seed, which could also grow sub-Eddington (e.g.\,\citealt{volonteri2010}). Current studies identify PopIII stars as main candidates of the progenitors of light seeds (e.g.\,\citealt{bond1984},\citealt{valiante2016}), while direct collapse of large primordial, low metallicity gas clouds produce $\sim10^{5-6} M_{\odot}$ seeds (e.g.\,\citealt{oh2002}, \citealt{begelman2006} \citealt{ferrara2014}). Alternatively, runaway collisions and stellar-dynamical interactions in dense primordial star clusters can form seeds with intermediate masses ($\sim 10^{3-4} M_{\odot}$; e.g.\,\citealt{devecchi2009}, \citealt{sakurai2017}). Another possibility to grow the observed black hole masses is a radiatively inefficient accretion scenario, which may allow for 100$\times$ higher mass accretion rates while remaining sub-Eddington, although this would require a large fraction of obscured quasars at high$-z$ (e.g.\,\citealt{davies2019}). Constraints from observational studies of black hole masses and accretion rates of sources at $z\gtrsim6$ are therefore fundamental to inform SMBHs' formation theories and models, and to position them in the context of their (co-)evolution with their host galaxies (e.g.\,\citealt{pensabene2020}, \citealt{neeleman2021}). 
   
   Currently, $\sim$100 measurements of $z\gtrsim5.8$ SMBH masses have been reported in the literature, based on ground-based NIR spectroscopic data with limited signal-to-noise ratios (SNR; e.g.\, \citealt{shen2019}, \citealt{yang2021}; see \citealt{fan2022} for a recent review). The backbone of such studies is the modelling of the region around the rest-frame UV \mgii\,2798\AA\, emission line, which can be used to derive black hole masses and accretion rates once virial equilibrium is assumed and taking advantage of scaling relations calibrated in the local universe (e.g.\,\citealt{vestergaard2009}). Another routinely used emission line to measure black hole masses is the \civ\,1549\,\AA\, (e.g.\,\citealt{vestergaard2006}), especially in cases in which the \mgii\, falls in or close to telluric absorption, even if one has to consider larger uncertainties due to intrinsic, non virial components of the \civ\, line, arising from winds/outflows (e.g.\,\citealt{coatman2017}). Although some studies suggest that $z\gtrsim 6$ quasars accrete at a rate comparable to that of a luminosity distribution-matched sample of quasars at $z\sim1-2$ (e.g.\,\citealt{mazzucchelli2017b}, \citealt{shen2019}), others observe a slight increase in Eddington ratio as a function of redshift (e.g.\,\citealt{yang2021}, \citealt{farina2022}). One of the main drawbacks of literature work so far is the low SNR of the considered data, which may introduce biases in the properties derived by the spectral fitting (e.g.\,\citealt{denney2016}), and could highly deteriorate the \civ\, or \mgii\, line modeling in case of strong absorption features.
   
   In this paper, we present measurements of bolometric luminosities, black hole masses and accretion rates from the modeling of the \mgii\, and \civ\, emission line regions for a sample of 42 luminous $z\sim6$ quasars from the enlarged XQR-30 survey (E-XQR-30). In particular, 30 objects were observed in the Legacy Survey of quasars at $z=$5.8--6.6 XQR-30 (\citealt{dodorico2023}; in this paper also referred to as ``{\it XQR30 Core}''), and 12 sources with similar properties and available X-Shooter observations with comparable SNR are obtained from the literature (in this paper also referred to as ``{\it XQR30 Extended}''). This is the first sample with such a high SNR ($\gtrsim 11 - 114$ per bin of 10 km s$^{-1}$; \citealt{dodorico2023}) optical/NIR spectra, which allows us for accurate modeling of their emission lines. Out of the total sample, black hole masses and accretion rates for 19 objects are reported for the first time. This work is organized as follows: in Section \ref{sec:sample}, we describe the sample and we briefly report the data reduction, while in Section \ref{sec:modeling} we report the spectral modeling; in Section \ref{sec:analysis} we present our measurements, we compare the black hole masses obtained via \civ\,and \mgii\,emission line modeling; we place our work in the context of quasars at lower redshift and we compare them with current measurements of high$-z$ quasars properties in the literature. We list our conclusions and outlook for future studies in Section \ref{sec:conclusion}. 
   Throughout the paper, magnitudes are reported in the AB system, and we use a flat cosmology with $H_0 = 70 \,\mbox{km\,s}^{-1}$\,Mpc$^{-1}$, $\Omega_M = 0.3$, and $\Omega_\Lambda = 0.7$.
   
%--------------------------------------------------------------------
\section{Sample and Data Reduction} \label{sec:sample}

   XQR-30 is an ESO Large Program (Program ID: 1103.A-0817(A); PI: D'Odorico) comprising high SNR ($\sim$11-41 in the continuum at rest-frame wavelength of 1285\,\AA) spectra for 30 high$-z$ quasars with the X-Shooter spectrograph \citep{vernet2011} at the VLT \citep{dodorico2023}. The quasars were selected to be observable from Paranal Observatory (Decl.$<$27deg), with redshift in the range $5.8\lesssim z \lesssim6.6$, and AB magnitude $J\lesssim19.8$ (20.0) for $z < 6.0$ ($ 6.0\lesssim z \lesssim$6.6). This survey aims at addressing several goals, from the characterization of the reionization process, to the study of absorbers along the line of sight and the early metal enrichment of the quasars' BLRs and circumgalactic medium (CGM). 
   We also consider 12 additional quasars with analogous luminosities and redshifts, and with X-Shooter spectra with comparable SNR ($\sim$17--114 at rest-frame 1285 \AA) available in the archive (data previously published in \citealt{beckerG2015}, \citealt{bosman2018}, \citealt{schindler2020}). The spectra for the entire sample were treated with a consistent methodology.
   
   Briefly, the data were reduced with a custom-made pipeline optimized for faint sources (\citealt{lopez2016}, \citealt{beckerG2019}). After standard reduction, the correction for telluric absorption is obtained using models created with ESO SKYCALC Cerro Paranal Advanced Sky Model (\citealt{noll2012}, \citealt{jones2013}). The relative flux calibration is measured with a static response function calculated from a standard star. The 1D stacked spectra of the VIS and NIR arms were combined using Astrocook (\citealt{cupani2018}, \citeyear{cupani2020}), and re-binned to a constant velocity step of 50 km s$^{-1}$.  Each quasar's spectrum was absolute-flux calibrated by scaling it to match the observed AB band magnitude in $J$ band \citep{dodorico2023}. For a full description of the sample and data reduction, see \cite{dodorico2023}.

\section{Modeling of the spectra} \label{sec:modeling}

   Quasars' rest-frame UV/optical spectra are characterized by a pseudo-continuum, due to different emission processes, and broad emission lines. To model these spectra, we followed the approach described in \cite{mazzucchelli2017b} and \cite{schindler2020}. Details of the spectral modeling will be presented in a forthcoming paper (Bischetti et al.\,2023, in prep.), while we summarize in the following the main spectral components:
   \begin{itemize}
       %------- Pseudo-continuum
       \item a \textit{power-law quasar continuum} emission. For the reddest quasars, we also included a second or third polynomial function to better fit the part of the spectrum blueward of the \civ\, (e.g.\,\citealt{shen2019}). To model this pseudo continuum, we consider regions of the spectra free of strong emission line features or of absorption due to the atmosphere (e.g.\,\citealt{schindler2020}).
       %------- Balmer pseudo continuum
       \item A \textit{Balmer pseudo continuum} ($f_{BC}$), as per the description from \cite{dietrich2003}:
       \begin{equation}
           f_{BC}(\lambda) = f_{BC,0} B_{\lambda} (\lambda,T_{e}) \left(1- e^{\tau_{\rm BE} (\lambda/\lambda_{\rm BE})^{3}} \right)
       \end{equation}
       with values for the electron temperature ($T_{\mathrm{e}}=15,000$ K) and optical depth ($\tau_{\mathrm{BE}}=1$) as used in other works (e.g.\,\citealt{derosa2014},\citealt{mazzucchelli2017b}). We impose the Balmer emission to 30\% of the pseudo-continuum contribution, as above described, at rest-frame $3646\,\mathrm{\AA}$ (e.g.\,\citealt{schindler2020}, \citealt{farina2022}). The Balmer pseudo-continuum and the power-law function are modelled at the same time.
       %------- FeII pseudo-continuum
       \item A \textit{\feii\, pseudo-continuum}, using the empirical template from \cite{vestergaard2001}, which is used to derive the \mgii\, emission line-based scaling relation to calculate black hole masses in Section \ref{sec:analysis}. The empirical template, shifted using an initial redshift measured from the \mgii\,emission line, is convolved with a Gaussian convolution kernel of different values depending on each spectrum.
       %------- Emission lines
       \item One (or more) Gaussian function(s) is used to model the broad emission lines, with a upper limit to the full-width-at-half-maximum (FWHM) $< 10000$ km s$^{-1}$, which prevents the model from using a very broad Gaussian function to model weak \feii\, emission not perfectly reproduced by the \cite{vestergaard2001} template.
   \end{itemize}
   
\begin{figure*}
   \centering
   \includegraphics[width=\hsize]{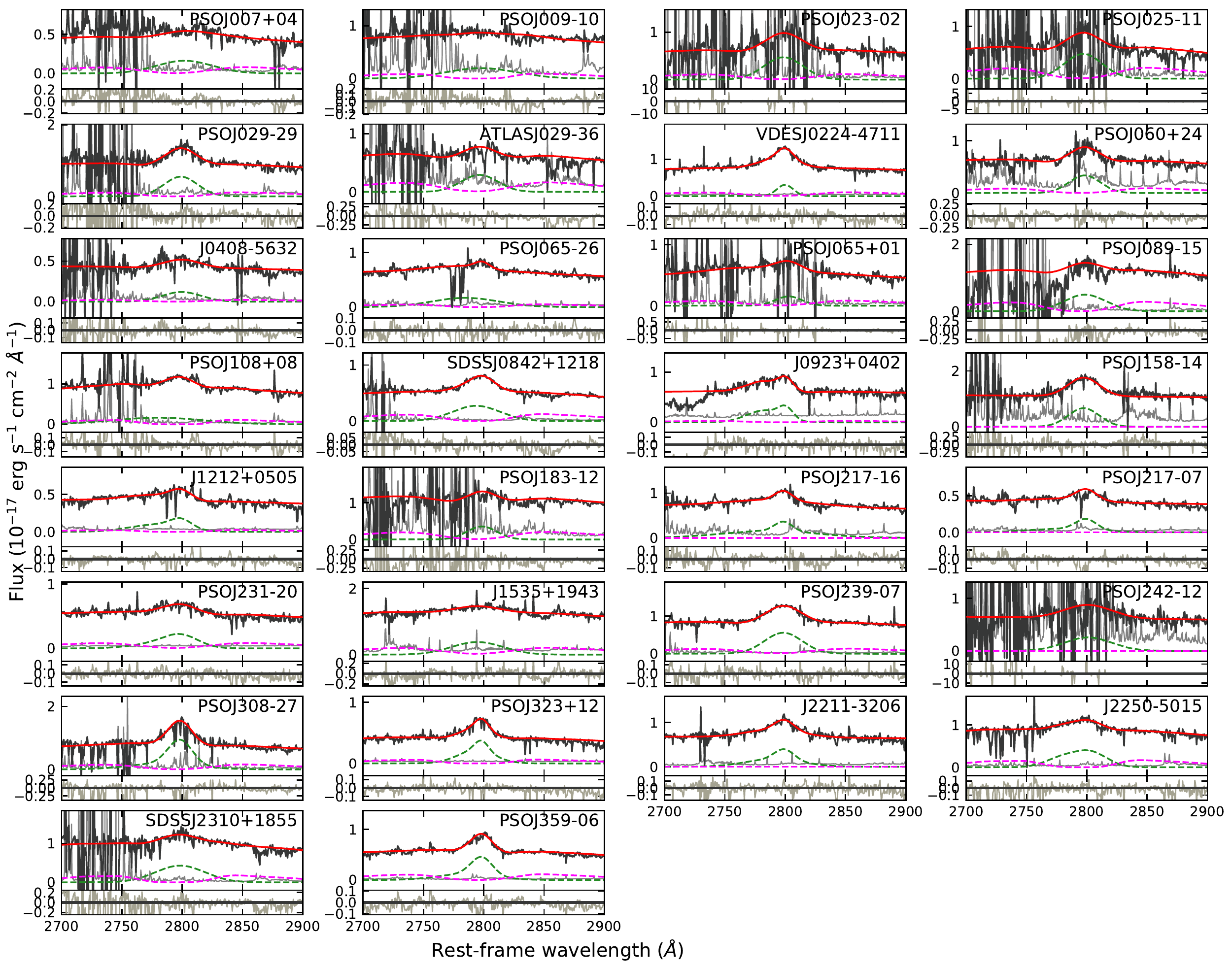}
   \caption{Spectral region centered on the \mgii\, emission line for the XQR-30 sample. The pseudo-continuum (power law + Balmer contribution + \feii\, empirical template) and multiple Gaussian lines fit are shown with dashed magenta and green lines, respectively, while the total fit is shown in red. We also show the residuals (in grey) in the lower panels. We note that in few cases the \mgii\, emission line falls partially (PSOJ007+04, PSOJ009-10, PSOJ183-12, PSOJ065+01) or fully (PSOJ023-02, PSOJ025-11, PSOJ242-12) in a region affected by telluric absorption. Despite the general high SNR of our spectra, given the (much) higher noise in these regions, the fit of these objects could be affected. }
              \label{fig:MgIIFit}%
    \end{figure*}

   Here, in case the emission lines were fit with more than one Gaussian function, we calculate the best values and uncertainties for the properties of the entire line following the method reported in \cite{schindler2020}. Briefly, for each line for which two or three Gaussian functions were used, we create N=1000 replication of each of the best Gaussian fits. For each replication, the mean and standard deviation of each fit are randomly drawn from Gaussian distributions, whose respective mean and standard deviation are fixed to the best fit value and associated uncertainty, respectively. Then, in each replication, we sum the single Gaussian functions and we calculate the total full-width-at-half-maximum (FWHM) as the distance between the two wavelengths where the flux is equal to half the maximum. All the 1000 FWHM values so obtained are distributed as a Gaussian function, and we consider as the best value and uncertainty for the final FWHM of the line the mean and sigma of such distribution. The final FWHM$\rm _{\mgii}$ and FWHM$\rm _{\civ}$ values are reported in Tab.\,\ref{table:sample_fit}. We show in Fig.\,\ref{fig:MgIIFit} the fit of the spectral region around the \mgii\, emission line. We note that in few cases, the \mgii\, emission line falls very closely (PSOJ007+04, PSOJ009-10, PSOJ183-12, PSOJ065+01) or within (PSOJ023-02, PSOJ025-11, PSOJ242-12) a region affected by strong telluric absorption. Therefore, also given the potential degeneracies of our spectral modeling with several components, the fit results and relative derived quantities (e.g.\,black hole masses, bolometric luminosities, Eddington ratios) should be taken with caution.

%-------------------------------------- MgII Spectral Fit -----------------

%--------------------------------------------------------------------
\section{Data Analysis} \label{sec:analysis}

   In this section, we derive the quasars' black hole masses, luminosities and accretion rates, relying on the fit of both the \civ\, and \mgii\, wavelength regions.
   
%-------------------------------------   
\subsection{Black hole masses and Eddington ratios calculation}  \label{sec:bh_calc}

   \textit{Mg II}: To derive black hole masses, we use the scaling relation provided by \cite{vestergaard2009}:
   %-----------------------
   \begin{equation}
      \mathrm{M_{BH,MgII}} = 10^{6.86} \, \left[ \frac{\mathrm{FWHM_{\mgii}}}{10^3 \, \mathrm{km\,s^{-1}}} \right]^{2} \left[ \frac{\lambda \mathrm{L}_{\lambda}\, (3000 \, \AA)}{10^{44} \, \mathrm{erg\,s^{-1}}} \right]^{0.5} \, M_{\odot}
   \end{equation}
   %-----------------------
   where $\lambda \mathrm{L}_{\lambda}(3000 \, \mathrm{\AA})$ is the monochromatic luminosity at rest-frame wavelength 3000 \AA. Systematic uncertainties in SMBH masses are estimated to be $\sim$0.55dex. \\[0.1cm]
   
   \noindent
   \textit{C IV}: In this case, we use the scaling relation from \cite{vestergaard2006}:
   %-----------------------
   \begin{equation} \label{eq:MBHCIV}
      \mathrm{M_{BH,CIV}} = 10^{6.66} \, \left[ \frac{\mathrm{FWHM_{\civ, corr}}} {10^3 \, \mathrm{km\,s^{-1}}} \right]^{2} \, \left[ \frac{\lambda \mathrm{L}_{\lambda} \, (1350 \, \AA)}{10^{44} \, \mathrm{erg\,s^{-1}}} \right]^{0.53} \, M_{\odot}
   \end{equation}
   %-----------------------
   where the $\lambda \mathrm{L}_{\lambda}(1350 \, \mathrm{\AA})$ is the monochromatic luminosity at rest-frame wavelength 1350 \AA\, and FWHM$\rm _{\civ, corr}$ is the corrected FWHM of the total \civ\, emission line. It is important to note that the \civ\, emission line profile is affected by the presence of an outflowing component (e.g.\,\citealt{richards2006a}, \citealt{meyer2019}). Therefore, we use the equation from \cite{coatman2017} to obtain the corrected value of the full-width half maximum:
   %-----------------------
   \begin{equation} \label{eq:FWHMCIVcorr}
      \mathrm{FWHM_{\civ, corr}} = \frac{\mathrm{FWHM_{\civ}}}{0.36 \times \frac{\civ\, \mathrm{Blueshift}}{10^{3} \mathrm{km s^{-1}}} + 0.61} 
   \end{equation}
   %-----------------------   
   where \civ\,Blueshift is the velocity difference between the \civ\, centroid and the quasars' systemic redshifts (obtained from the \mgii \, or from the \cii \, emission line, when available, as reported in \citealt{dodorico2023}, and in Tab.\,\ref{table:sample_fit}), in units of km s$^{-1}$. In this case, the uncertainties measured on the \cite{vestergaard2006} relation is $\sim$0.40 dex, while it is estimated to be reduced to $\sim$0.24 dex with the correction by \cite{coatman2017}. We calculated both $\mathrm{L}_{\lambda}(1350 \mathrm{\AA})$ and $\mathrm{L}_{\lambda}(3000 \mathrm{\AA})$ using the value of the fluxes at 1350 \AA\, and 3000 \AA\, from the power-law fit.\\[0.1cm]

   \noindent
   From the black hole mass measurements, we can calculate the Eddington luminosity as:
   %-----------------------
   \begin{equation}
      L_{\mathrm{Edd, \civ / \mgii}} = 1.3 \times 10^{38} \, \left( \frac{\mathrm{M_{BH, \civ / \mgii}}}{M_{\odot}} \right) \, \mathrm{erg\,s^{-1}}
   \end{equation}
   %-----------------------
   We also compute the bolometric luminosity ($L_{\mathrm{bol}}$) using the bolometric correction presented by \cite{richards2006a}:
   %-----------------------
   \begin{equation}
      L_{\mathrm{bol}} = 5.15 \lambda \, L_{\lambda} (3000 \mathrm{\AA}) \, \mathrm{erg\,s^{-1}}
   \end{equation}
   %-----------------------
   We note that it has been discussed that such bolometric correction might be overestimated for highly luminous quasars (e.g.\,\citealt{trakhtenbro2012}). Nevertheless, we decide to use it for consistency with several works in the literature (e.g.\,\citealt{mazzucchelli2017b}, \citealt{yang2021}, \citealt{farina2022}).
   From the $L_{\mathrm{Edd}}$ and $L_{\mathrm{bol}}$ values we can derive the corresponding Eddington ratios $\lambda_{\mathrm{Edd, \civ}} = L_{\mathrm{bol}} / L_{\mathrm{Edd, \civ}}$ and $\lambda_{\mathrm{Edd, \mgii}}  = L_{\mathrm{bol}} / L_{\mathrm{Edd, \mgii}}$.
   We report the values of the monochromatic and bolometric luminosities, black hole masses and Eddington ratios in Tab. \ref{table:sample_fit}.

%-------------------------------------   
\subsection{\civ\, vs \mgii\, black hole masses comparison} 

\begin{figure}
   \centering
   \includegraphics[width=0.75\hsize]{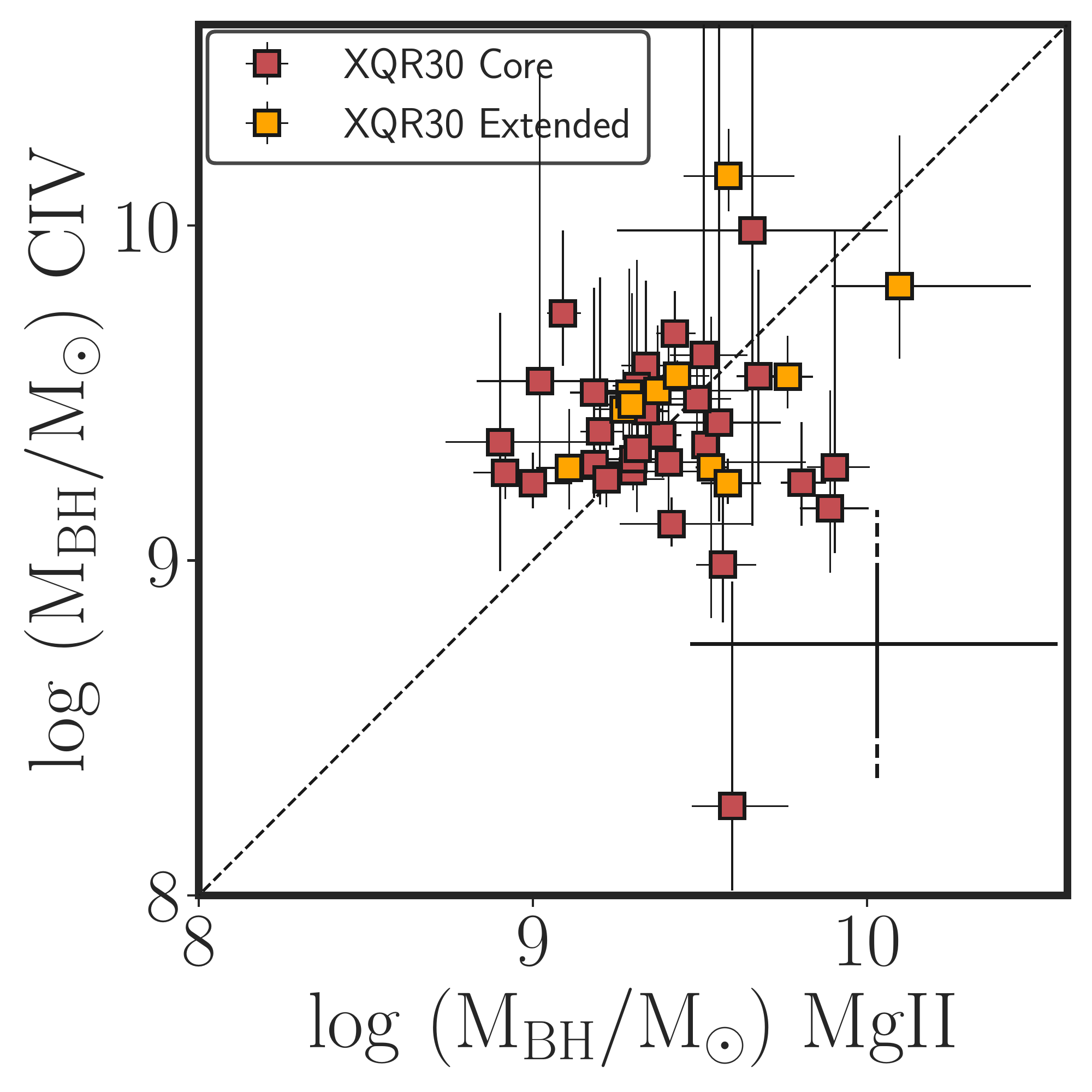}
   \caption{Comparison between \civ- and \mgii-based black hole masses. Typical uncertainties due to the scatter in the relations used are shown with a black cross in the right-bottom corner. Uncertainties on the uncorrected \cite{vestergaard2006} scaling relations ($\sim$ 0.40 dex) are shown with dashed lines. }
              \label{fig:MBHCIVMgII}%
    \end{figure}

   We compare the black hole masses measured from the \civ\, and \mgii\,emission line regions modeling in Fig.\,\ref{fig:MBHCIVMgII}. The two measurements are approximately in the same range, although the \civ\, based values are slightly higher than the \mgii\, ones, with a mean ratio of $M_{\rm BH, \civ} / M_{\rm BH, \mgii}\sim$1.3. On the other hand, as discussed in \cite{farina2022}, we also notice that the \civ\, modeling tends to underestimate the values of black hole masses for higher $M_{\rm BH, \mgii}$ values, i.e.\,for higher FWHM$_{\rm \mgii}$ values. Finally, we also notice that a high fraction of broad absorption line (BAL) quasars have been recovered in the XQR-30 sample ($\sim 50 \%$; \citealt{bischetti2022}). Even though BAL features may complicate the fit of the \civ\, emission line region, the high SNR of our spectra still permits a good modeling of the line in the majority of the cases. The quasar PSO~J065+01 stands out as a particular outlier, with a \civ-based mass lower than that recovered from the \mgii\, line by a factor of $\sim$1.3dex. This is due to the very peculiar shape of the quasar spectrum, and the very low SNR of the \civ\, emission line.
   
   If one considers the \mgii\,line-derived BH masses as the reference values, we can estimate the mean (median) of the dispersion of the \civ\,line-derived values to be {\bf 0.28 (0.21)} dex. If one excludes PSOJ065+01, one obtains a mean (median) dispersion of 0.25 (0.21) dex. All these values are lower than the dispersion of the \civ\,-based scaling relation to obtain black hole masses (see Eq.\,\ref{eq:MBHCIV}), expected to be $\sim$0.40dex from the \cite{vestergaard2006} relation, and are consistent with the $\sim$0.24dex scattering expected when considering the \cite{coatman2017} correction. A larger black hole masses range would be needed to fully understand how the \civ-\, and \mgii-based black hole masses estimates compare. 
   
%-------------------------------------- MgII vs CIV comparison -----------------

%-------------------------------------- Black Hole vs Lbol -----------------
   \begin{figure}
   \centering
   \includegraphics[width=\hsize]{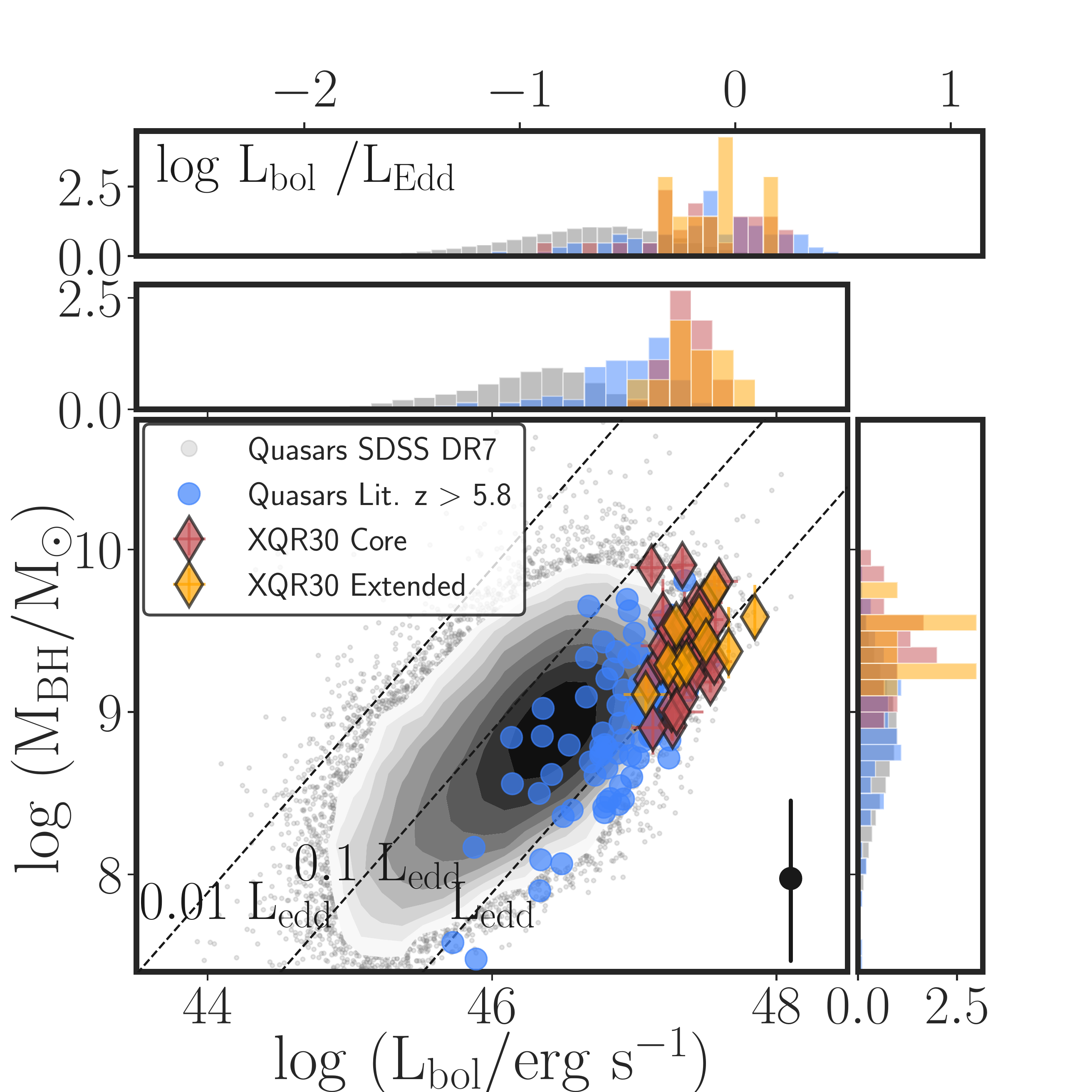}
   \caption{Black hole masses versus bolometric luminosities. We show quasars at $0.35 < z < 2.25$ from the SDSS DR7 survey (black contours and grey points), and a sample of $z > 5.8$ quasars from the literature (blue points, see Section \ref{sec:MBHSampComp} for references). We report the newly calculated values for the core XQR-30 quasars in red diamonds, and the additional 12 high-z sources in golden diamonds. Typical systematic uncertainty on black hole masses, due to the scaling relations used ($\sim$0.55dex), is shown in the bottom right corner. Distributions of bolometric luminosities, black hole masses and Eddington ratios for all the above described samples are also reported.}
              \label{FigMBHLbol}%
    \end{figure}

%-------------------------------------   

\subsection{Comparison with the literature and lower redshift samples}  \label{sec:MBHSampComp}

   We place our black hole mass and Eddington ratio measurements, based on the \mgii\, modeling, in the context of the literature (e.g.\,\citealt{yang2021}, \citealt{farina2022}). For simplicity, in the following section, we will denominate these quantities as $M_{\rm BH}$ and $\lambda_{\mathrm{BH}}$.
   
   To consider sources at low redshift, we take the SDSS Data Release 7 (DR7, \citealt{richards2011}) quasar catalog. We select objects with: {\it a)} redshift between $0.35 < z < 2.25$, i.e.\, with the \mgii\, emission line region recovered in the SDSS wavelength range; {\it b)} valid values for $L_{\lambda}(3000)$; {\it c)} broad \mgii\, emission line (FWHM$_{\mgii} > 1000.$ km s$^{-1}$), with spectra of good quality and hence reliable fit (FWHM$_{\mgii} > 2 \, \mathrm{ERR\_FWHM_{\mgii}}$ and EW$_{\mgii} > 2 \, \mathrm{ERR\_EW_{\mgii}}$). The sample so-obtained is of 77,824 quasars.
   Regarding the high-redshift sample, we consider quasars at $z>5.8$ with NIR spectra observed in the literature (\citealt{willott2010a}, \citealt{derosa2011}, \citealt{mazzucchelli2017b}, \citealt{chehade2018}, \citealt{shen2019}, \citealt{matsuoka2019}, \citealt{reed2019}, \citealt{onoue2019}, \citealt{andika2020}, \citealt{eilers2020a}, \citealt{schindler2020}, \citealt{banados2021}, \citealt{yang2021}, \citealt{farina2022}; see \citealt{fan2022} for a review). We obtained 114 sources, out of which 23 are also part of the E-XQR-30 sample. We report in Section \ref{sec:appA} a comparison between the values obtained in this work and in the literature, showing an overall agreement. For the remaining part of the analysis presented below, we consider for these quasars the values of $M_{\rm BH}$, $L_{\rm bol}$ and $\lambda_{\mathrm{BH}}$ newly derived here. In the following comparison, we exclude the quasar J0100+2802, which is a strong outlier in bolometric luminosity and does not have comparable counterparts in the SDSS survey, and the quasar J0439+1634, which is gravitationally lensed (\citealt{fan2019}). Hence, the total sample of high redshift quasars considered here (literature + E-XQR-30) is of 133 sources.
   For both the high-z quasars from the literature and the low-redshift objects from SDSS, we calculate the values of black hole masses, bolometric luminosities and Eddington ratios in a consistent way with the sample presented in this paper (see Sec.\, \ref{sec:bh_calc}). We show in Fig.\, \ref{FigMBHLbol} the black hole masses and bolometric luminosites values for the E-XQR-30 sample and for these comparison samples. As already expected from the sample selection, we see that the E-XQR-30 sample occupies the parameter space at the highest luminosities, with a mean (median) $L_{\mathrm{bol}}$ value of 2.3 (2.2) $\times 10^{47}$ erg s$^{-1}$. The mean (median) $M_{\mathrm{BH}}$ values are of 2.9 (2.4) $\times 10^{9}$ $M_{\odot}$, and mean (median) $\lambda_{\mathrm{BH}}$ values of 0.84 (0.72).
   
   In general, caution should be taken when comparing different quasars sample. First of all, it is extremely difficult to define the completeness of the high-z sample, due to the heterogeneous selection criteria of the different sub-samples (e.g.\,\citealt{banados2016}, \citealt{jiang2016}, \citealt{matsuoka2019}). This can insert systematic biases in our comparison, considering that even well defined samples can be biased (for instance a positive luminosity-dependent bias of measured black hole masses has been found; e.g.\, \citealt{shen2012}, \citealt{kelly2013}, \citealt{wu2022}). Secondly, we would like to highlight that, given that we derived $L_{\mathrm{bol}}$ using the bolometric correction from \cite{richards2006a}, these values are a reflection of the quasars UV luminosities. Hence, when we match samples by bolometric luminosity distribution (see below), we are effectively considering the intrinsic UV luminosity distribution. Finally, our cut of FWHM$_{\mgii} > 1000$ km s$^{-1}$ in the low-redshift quasars selection, despite being generally considered in literature for defining broad emission lines objects (e.g.\, \citealt{padovani2017}), can insert a bias against slightly lower FWHM values, which in return affects BHs with lower masses. Keeping in mind these cautions, we still decide to compare the high-redshift quasars sample and the quasars at lower redshifts, in order to test for any redshift evolution in the black hole and Eddington ratio distributions.
   
   We first consider only the sources in the E-XQR30 sample, then we utilize all high$-z$ quasars (from this work + the literature). In order to obtain a consistent comparison, we select a subsample of quasars from the SDSS matching the bolometric luminosity distribution of the sample of quasars at high$-z$. In practice, we select sources at low$-z$ in a range of $\pm$0.01 log$L_{\mathrm{bol}}$ for each high$-z$ quasar, and we consider their respective $M_{\mathrm{BH}}$ and $\lambda_{\mathrm{BH}}$: we repeat this trial 1000 times. The mean (median) of the black hole masses values of the low$-z$ quasar sample matched with the E-XQR30 sources are 3.2 (2.6) $\times 10^{9}$ $M_{\odot}$. On the other hand, the mean (median) $\lambda_{\mathrm{BH}}$ values are 0.86 (0.69). We note that these values are comparable with those obtained for the E-XQR30 objects. We also performed a Kolmogorov-Smirnov (KS) test in order to assess if the bolometric-luminosity matched low$-z$ and the E-XQR30 samples are consistent to be drawn from the same underlying population. We obtain a p value of 0.41 / 0.27 for the $M_{\mathrm{BH}}$ / $\lambda_{\mathrm{BH}}$ distribution, rejecting the hypothesis that these two samples are not drawn from the same population.
   
\begin{figure*}
   \centering
   \includegraphics[width=0.9\hsize]{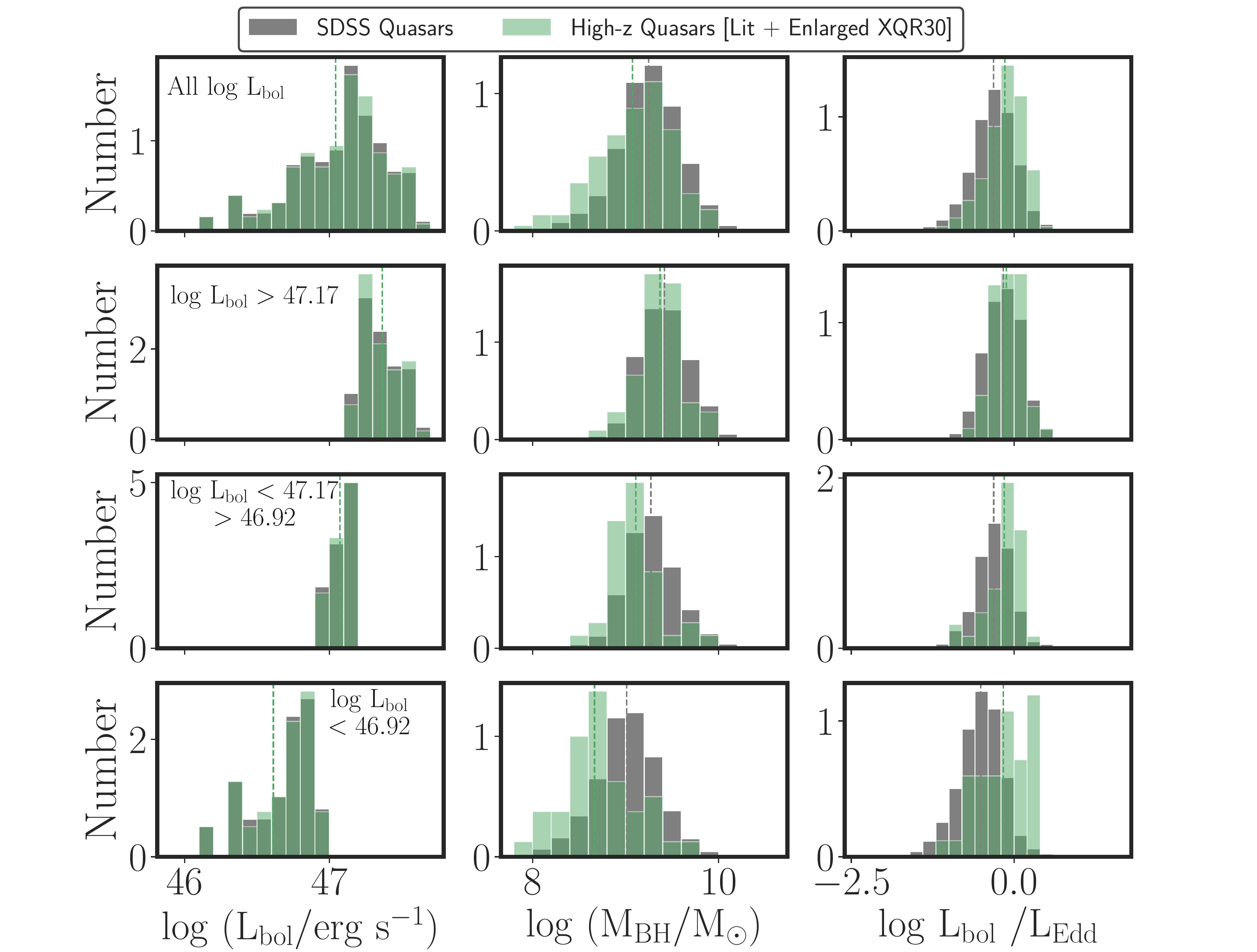}
   \caption{Bolometric Luminosities ({\it left}), black hole masses ({\it middle}) and Eddington ratios ({\it right}) distributions for all the quasars at high$-z$ (literature + enlarged XQR-30) in green, and for a bolometric luminosity matched sample at low$-z$. Mean values for the quantities, both at low- and high-redshift, are shown with dashed lines, in grey and green, respectively. In the {\it top} row we consider all the luminosity distribution of high$-z$ quasars, while from the {\it second} to the {\it last} row we present the samples divided by luminosity ranges, each containing the same number of high$-z$ quasars, from high to medium to low luminosities. See Section \ref{sec:MBHSampComp} for further details on the samples compilation and matching. The Eddington ratio distribution at high$-z$ is shifted to marginally higher values than at lower$-z$, for the entire luminosity case, and for each luminosity range. We note that with decreasing luminosity, the separation between the high and low redshift sample is increasing, albeit with larger dispersions.}
              \label{fig:MBHLbolHisto}%
    \end{figure*}

   We now consider the entire high-z sample (this work + literature; 133 objects). In order to test how our comparison relies on the intrinsic high$-$redshift quasars luminosity distribution, we repeat the same comparison as above in different ranges of luminosity, following the approach from \cite{farina2022}. We consider three luminosity ranges each containing the same number of quasars: high luminosity (log $L_{\mathrm{bol}} > 47.17$ erg s$^{-1}$), medium luminosity ( $46.92 < {\rm log}\, L_{\mathrm{bol}} < 47.17$ erg s$^{-1}$) and low luminosity (log $L_{\mathrm{bol}} < 46.92$ erg s$^{-1}$). Results are reported in Fig.\, \ref{fig:MBHLbolHisto}. The mean, median, and standard deviation values for $L_{\mathrm{bol}}$, $M_{\rm BH}$ and $\lambda_{\mathrm{BH}}$, for the entire luminosity sample, and for each range of luminosity, for the low$-$ and high$-$z samples are listed in Tab.\,\ref{table:mean}. We repeated the KS test for this sample, and for all the luminosity ranges. The resulting p-values are also reported in Tab.\,\ref{table:mean}.
   
   We note that the bolometric luminosity distribution in the low$-z$ sample is constructed to be consistent with that at high$-z$. This is also reflected in the corresponding p-values of 1 for all the cases (see Tab.\,\ref{table:mean}). We find that the mean and median black hole mass/Eddington ratio values are lower/higher in the high-redshift sample with respect to the low$-z$ one, considering all luminosities, and in every luminosity range. However, we note that these differences with redshifts are more significant at lower luminosities, with a difference between the mean Eddington ratio at high$-$ and low$-z$ of $\sim$0.38 in the low luminosity range, higher with respect to what is observed for luminous objects ($\sim$0.03). We can also notice that the distributions of M$_{\rm BH}$ and $\lambda_{\mathrm{BH}}$ have slightly larger dispersions at lower luminosities. E.g.\,the standard deviation for the high$-z$ black hole masses (Eddington ratios) is 0.45 (0.38) dex in the low luminosity range, with respect to a standard deviation of 0.25 (0.24) dex in the high luminosity range. These trends can also be reflected in the results of the KS test. Indeed, in the high luminosity range, the p-values obtained by comparing the black hole masses and Eddington ratios distributions at high$-$ and low$-z$ are relatively high (0.47 and 0.32, respectively). Conversely, in the medium and low luminosity ranges, and when considering all luminosities, we recover low p-values ($< 10^{-4}$), rejecting the hypothesis that these quantities are drawn from the same underlying distribution.
   
   In summary, our analysis suggests that quasars at high redshift accrete slightly faster than those in a bolometric luminosity distribution matched sample at $z\sim1.5$, assuming the same mean radiative efficiency. This trend increases for the faintest objects discovered, albeit with a larger dispersion. In other words, at high-redshift we observe that the most luminous quasars are powered by less massive SMBHs, accreting at slightly higher rates compared to a luminosity-matched sample at high$-z$. This result is in contrast with respect to previous works (e.g.\,\citealt{mazzucchelli2017b}, \citealt{shen2019}), which did not recover a change in the mean Eddington ratio value with $z$, and with expectations from the consistency between composite spectra of quasars at $z\gtrsim6$ and at lower$-z$ (e.g.\,\citealt{shen2019}, \citealt{yang2021}). On the other hand, our outcome is in agreement with the recent results by \cite{yang2021} and \cite{farina2022}, which observed a similar increase in the mean value of the Eddington ratio as measured in this work. Also, one can notice that the composite spectra obtained in the literature are usually focused on the higher luminosity quasars, both at high- and low$-z$ (e.g.\,\citealt{vandenberk2001}, \citealt{selsing2016}), where the changes between the two samples are less apparent (see Fig.\,\ref{fig:MBHLbolHisto} and Tab.\,\ref{table:mean}).

%-------------------------------------- Histo comparison -----------------
    
%-------------------------------------   
\subsection{Comparison between XQR-30 and literature values} \label{sec:appA}
\begin{figure}
   \centering
   \includegraphics[width=0.7\hsize]{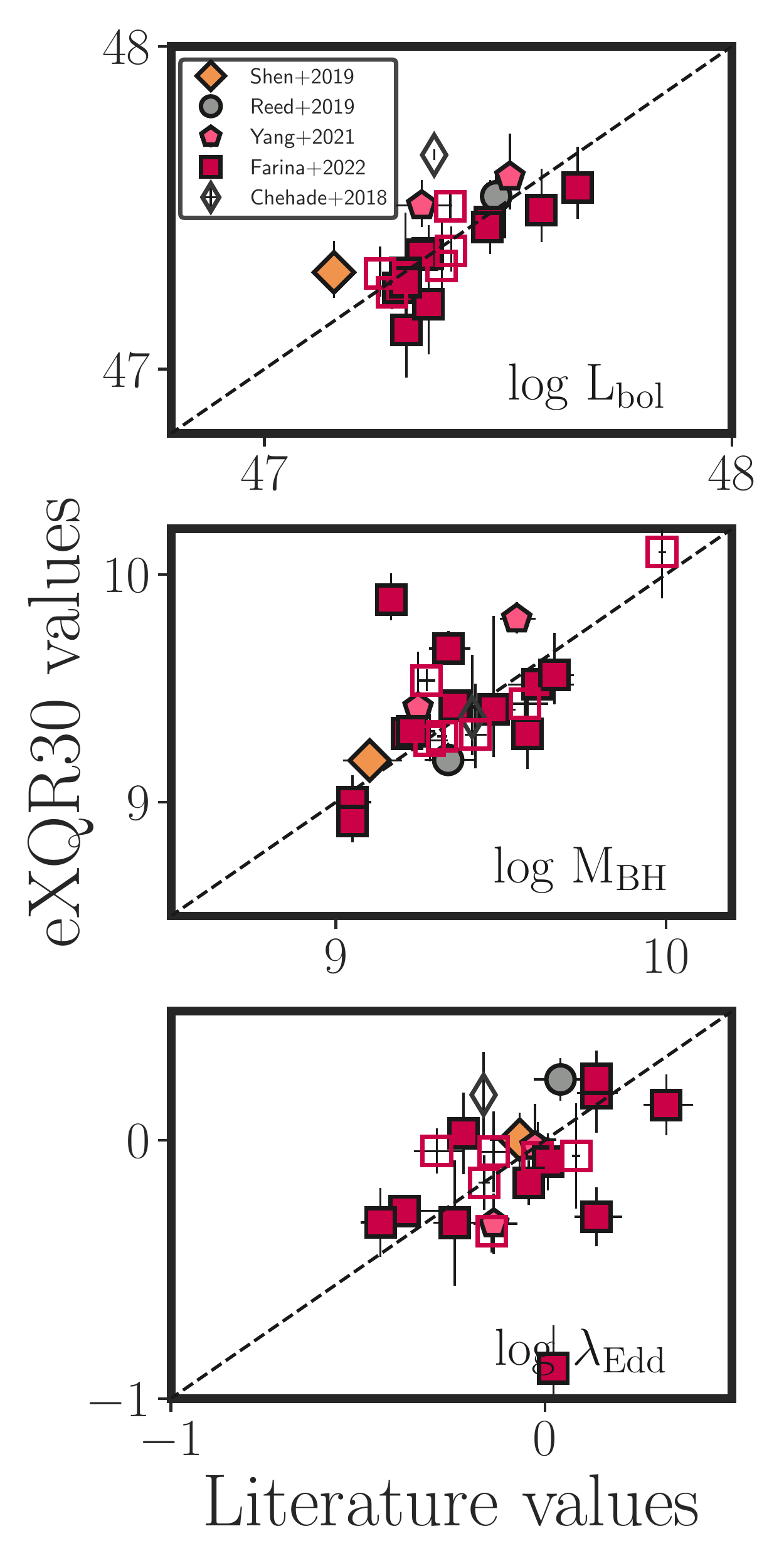}
   \caption{Comparison between bolometric luminosities ({\it upper}), black hole masses ({\it central}) and Eddington ratios ({\it lower panel}) for the quasars in the enlarged XQR-30 sample that were previously observed and studied in the literature: black diamond (\citealt{chehade2018}), yellow diamond (\citealt{shen2019}), grey circle (\citealt{reed2019}), pink pentagon (\citealt{yang2021}) and red squares (\citealt{farina2022}). The dashed black line denotes the one-to-one relation. The filled symbols represent data of sources from the core XQR-30 survey, while empty ones show data from the literature sample.}
    \label{fig:MBHLbollit}
    \end{figure}

We note that 23 objects reported in the E-XQR-30 sample already have observations reported in the literature. In case of sources observed by more than one study, we consider only the most recent measurement: J0142-3327 (\citealt{chehade2018}); PSO060+24 (\citealt{shen2019}); J0224-4711 (\citealt{reed2019}); J0923+0402, J1535+1943 (\citealt{yang2021}); PSO239-07, J2211-3206, J0842+1218, PSO231-20, PSO158-14, PSO007+04, PSO065-26, PSO183+05, J2310+1855, PSOJ359-06, PSOJ323+12, J1319+0950, PSO036+03, J1030+0524, J1306+0356, J1509-1749, J0100+2802 (\citealt{farina2022}). The quasar J0439+1634 was also observed by \cite{yang2021}, but given that this source is lensed (\citealt{fan2019}), we decide to not include it in this comparison. We show in Fig.\,\ref{fig:MBHLbollit} the comparison of the bolometric luminosities, black hole masses and Eddington ratios (from the \mgii\, emission line model), all calculated with a consistent method (see Section \ref{sec:bh_calc}). In general, we note that there are no recovered systemic trends between the quantities derived here and those from the literature. The major outlier reported here is PSOJ007+04, whose \mgii\, emission line was fitted in this work with a very broad Gaussian (see Tab.\,\ref{table:sample_fit} and Fig.\,\ref{fig:MgIIFit}). This resulted in a very large black hole mass and a low Eddington ratio, differently than what presented in \cite{farina2022}. This is due to the fact that the \mgii\,line in this quasar is very close to a telluric absorption, hence its modelling should be considered with caution (see Fig.\,\ref{fig:MgIIFit} and Tab.\,\ref{table:sample_fit}).

%-------------------------------------- Comparison literature-----------------
    
%-------------------------------------   
\section{Conclusions} \label{sec:conclusion}

   The E-XQR-30 sample provides us with a unique opportunity to study quasars at high$-z$ with spectra of exquisite quality and high signal-to-noise ratio. Here, we calculate their bolometric luminosities via the monochromatic luminosity at rest frame 3000 \AA. We obtain black hole mass values by modeling the \civ\, and \mgii\, emission line regions, using scaling relation calibrated in the local universe (\citealt{vestergaard2006}, \citealt{vestergaard2009}). In particular, we account for the non-virial component of the \civ\, emission line, due to outflows/wings, utilizing the correction from \cite{coatman2017}. 
   We observe that, in our sample, the black hole mass values obtained with the two emission lines cover the same parameter space (see Fig.\,\ref{fig:MBHCIVMgII}).
   Assuming that the \mgii-based estimates are the more reliable, the scatter of the \civ-based measurements is lower than that measured around the \cite{vestergaard2006} scaling relation ($\sim$0.4\,dex), and it is consistent/slightly larger than what is expected after the \cite{coatman2017} correction ($\sim$0.24\,dex). 
   
   We compare the values measured from the E-XQR-30 objects with those of other quasars at $z>5.8$ obtained from the literature, and from a sample of quasars at $0.35<z<2.25$ from the SDSS DR7 survey (see Fig.\,\ref{FigMBHLbol}). We consider a comparison sub-sample of quasars at low$-z$, matched to the entire high$-z$ quasars' bolometric luminosity distribution. The high$-$redshift Eddington ratio distribution is slightly higher with respect to the matched low$-z$ sample (i.e.\,with a difference between the mean $\lambda_{\mathrm{BH}}$ values of $\sim$0.27; see Fig.\,\ref{fig:MBHLbolHisto} and Tab.\,\ref{table:mean}). We repeat this comparison considering sub-samples of high$-$ and low$-z$ quasars in different bolometric luminosity ranges, and noted that this increase in the mean Eddington ratio at higher redshifts is present in every luminosity range, and is more marked at lower luminosities. This suggests that quasars at $z\gtrsim6$ accrete marginally faster than at $z\sim1$, as suggested by other recent works in the literature (e.g.\,\citealt{yang2021}, \citealt{farina2022}).
   
   In the future, it will be crucial to explore the properties of quasars over larger ranges of luminosity and black hole masses, thanks also to the discoveries that will be enabled by future surveys such as the Legacy Survey of Space and Time (LSST) at Vera Rubin Observatory (e.g.\,\citealt{ivezic2014}, \citeyear{ivezic2019}) and the Euclid survey (e.g.\,\citealt{euclidcollaboration2019}). On the other hand, it will also be important to measure black hole masses from other emission lines, e.g.\,H$\beta$, which are directly related to the reverberation mapping studies at low$-z$, and test whether there are any systematic difference with values obtained from the \civ\, or \mgii\, line modeling (e.g.\,\citealt{homayouni2020}). The James Webb Telescope will be key in pursuing such studies (e.g.\,\citealt{eilers2023}, \citealt{larson2023}, \citealt{yang2023}, \citealt{maiolino2023}).   

%----------------------- Fit of the lines
\begin{sidewaystable*}
\footnotesize 
\centering 
\caption{Enlarged XQR-30 Sample: Redshift derived from the \mgii\,emission line modeling or, when available, from the \cii\, emission line (see \citealt{dodorico2023}); \civ\, and \mgii\, full-width at half maximum; \civ \, blueshift (used in equation \ref{eq:FWHMCIVcorr}); monochromatic luminosities at rest-frame 1350 \AA\, and 3000 \AA; bolometric luminosities; black hole masses and Eddington ratio values, derived from both the \civ\, and \mgii\, emission lines.}.             
\label{table:sample_fit}      
\begin{tabular}{l c c c c c c c c c c c c}     % 7 columns 
\hline\hline
%----------------------------------------------------------------------------
Name & Redshift & $\mathrm{FWHM_{C\,IV}}$ & $\mathrm{FWHM_{Mg\,II}}$ & \civ\, $\mathrm{Blueshift}$ & log $\lambda L_{1350}$ & log $\lambda L_{3000}$ & log $L_{\mathrm{bol}}$ & log $M_{\mathrm{BH,\,C\,IV}}$ & log $M_{\mathrm{BH,\,Mg\,II}}$ & $\lambda_{\mathrm{Edd,C\,IV}}$ & $\lambda_{\mathrm{Edd,Mg\,II}}$ \\  
& & [km s$^{-1}$] &  [km s$^{-1}$] & [km s$^{-1}$] & [erg s$^{-1}$] & [erg s$^{-1}$]  & [erg s$^{-1}$] & [M$_{\odot}$] & [M$_{\odot}$] \\[0.1cm]
PSOJ007+04$^{\rm a)}$ & 6.0015$^{\rm \dagger}$ & 6169$\pm$1057 & 8176$\pm$473 & 2778$\pm$601 & 46.56$^{+0.05}_{-0.06}$ & 46.41$^{+0.15}_{-0.22}$ & 47.12$^{+0.15}_{-0.22}$ & 9.16$^{+0.19}_{-0.35}$ & 9.89$^{+0.09}_{-0.11}$ & 0.71$\pm$0.49 & 0.13$\pm$0.06\\ 
PSOJ009-10$^{\rm \ddagger a)}$ & 6.004$^{\rm \dagger}$ & 8025$\pm$2963 & 7336$\pm$697 & 2956$\pm$456 & 46.42$^{+0.06}_{-0.06}$ & 46.63$^{+0.08}_{-0.09}$ & 47.34$^{+0.08}_{-0.09}$ & 9.28$^{+0.26}_{-0.71}$ & 9.9$^{+0.08}_{-0.1}$ & 0.88$\pm$0.73 & 0.21$\pm$0.06\\ 
PSOJ023-02$^{\rm \ddagger a)}$ & 5.90 & 4806$\pm$486 & 4065$\pm$201 & 832$\pm$112 & 46.39$^{+0.04}_{-0.05}$ & 46.62$^{+0.06}_{-0.07}$ & 47.33$^{+0.06}_{-0.07}$ & 9.37$^{+0.13}_{-0.18}$ & 9.39$^{+0.05}_{-0.06}$ & 0.7$\pm$0.26 & 0.68$\pm$0.12\\ 
PSOJ025-11$^{\rm a)}$ & 5.85 & 5283$\pm$1043 & 3976$\pm$196 & 1316$\pm$286 & 46.68$^{+0.03}_{-0.04}$ & 46.56$^{+0.09}_{-0.11}$ & 47.27$^{+0.09}_{-0.11}$ & 9.45$^{+0.2}_{-0.39}$ & 9.34$^{+0.06}_{-0.07}$ & 0.51$\pm$0.32 & 0.66$\pm$0.18\\ 
PSOJ029-29 & 5.984 & 6886$\pm$1190 & 3501$\pm$292 & 2063$\pm$285 & 46.88$^{+0.02}_{-0.03}$ & 46.78$^{+0.07}_{-0.08}$ & 47.49$^{+0.07}_{-0.08}$ & 9.58$^{+0.16}_{-0.25}$ & 9.34$^{+0.07}_{-0.09}$ & 0.63$\pm$0.3 & 1.1$\pm$0.27\\ 
ATLASJ029-36 & 6.02 & 5808$\pm$1517 & 3753$\pm$266 & 1924$\pm$377 & 46.73$^{+0.01}_{-0.01}$ & 46.38$^{+0.03}_{-0.03}$ & 47.1$^{+0.03}_{-0.03}$ & 9.39$^{+0.22}_{-0.46}$ & 9.2$^{+0.06}_{-0.07}$ & 0.4$\pm$0.26 & 0.6$\pm$0.1\\ 
VDESJ0224-4711 & 6.526 & 5378$\pm$207 & 2863$\pm$206 & 1808$\pm$42 & 46.61$^{+0.04}_{-0.05}$ & 46.82$^{+0.06}_{-0.06}$ & 47.54$^{+0.06}_{-0.06}$ & 9.29$^{+0.04}_{-0.05}$ & 9.19$^{+0.06}_{-0.08}$ & 1.36$\pm$0.23 & 1.72$\pm$0.36\\ 
PSOJ060+24 & 6.18 & 4522$\pm$625 & 3270$\pm$249 & 633$\pm$323 & 46.58$^{+0.04}_{-0.04}$ & 46.59$^{+0.08}_{-0.1}$ & 47.3$^{+0.08}_{-0.1}$ & 9.5$^{+0.31}_{-0.31}$ & 9.18$^{+0.07}_{-0.09}$ & 0.48$\pm$0.52 & 1.01$\pm$0.27\\ 
J0408-5632$^{\rm \ddagger}$ & 6.0345 & 6943$\pm$355 & 4057$\pm$323 & 2411$\pm$155 & 46.55$^{+0.03}_{-0.04}$ & 46.48$^{+0.08}_{-0.1}$ & 47.19$^{+0.08}_{-0.1}$ & 9.33$^{+0.07}_{-0.08}$ & 9.31$^{+0.08}_{-0.09}$ & 0.55$\pm$0.15 & 0.57$\pm$0.16\\ 
PSOJ065-26 & 6.1871$^{\rm \dagger}$ & 8866$\pm$2638 & 4878$\pm$836 & 3895$\pm$1503 & 46.83$^{+0.01}_{-0.01}$ & 46.64$^{+0.04}_{-0.04}$ & 47.35$^{+0.04}_{-0.04}$ & 9.41$^{+0.3}_{-1.6}$ & 9.56$^{+0.13}_{-0.18}$ & 0.67$\pm$0.66 & 0.48$\pm$0.17\\ 
PSOJ065+01$^{\rm \ddagger ab)}$ & 5.79 & 2389$\pm$878 & 5569$\pm$849 & 2682$\pm$368 & 46.4$^{+0.05}_{-0.05}$ & 46.49$^{+0.08}_{-0.1}$ & 47.2$^{+0.08}_{-0.1}$ & 8.27$^{+0.25}_{-0.67}$ & 9.6$^{+0.12}_{-0.17}$ & 6.63$\pm$5.37 & 0.31$\pm$0.12\\ 
PSOJ089-15$^{\rm \ddagger}$ & 5.957 & 3345$\pm$707 & 4365$\pm$425 & 1385$\pm$164 & 46.6$^{+0.04}_{-0.05}$ & 46.86$^{+0.05}_{-0.06}$ & 47.57$^{+0.05}_{-0.06}$ & 8.99$^{+0.17}_{-0.29}$ & 9.57$^{+0.08}_{-0.1}$ & 2.94$\pm$1.48 & 0.77$\pm$0.18\\ 
PSOJ108+08 & 5.9485 & 8164$\pm$628 & 4247$\pm$346 & 3109$\pm$281 & 46.84$^{+0.04}_{-0.04}$ & 46.75$^{+0.1}_{-0.13}$ & 47.46$^{+0.1}_{-0.13}$ & 9.48$^{+0.09}_{-0.12}$ & 9.49$^{+0.08}_{-0.1}$ & 0.74$\pm$0.26 & 0.72$\pm$0.24\\ 
SDSSJ0842+1218$^{\rm \ddagger}$ & 6.0754$^{\rm \dagger}$ & 6041$\pm$270 & 3854$\pm$337 & 2078$\pm$37 & 46.58$^{+0.02}_{-0.02}$ & 46.54$^{+0.05}_{-0.05}$ & 47.25$^{+0.05}_{-0.05}$ & 9.3$^{+0.04}_{-0.05}$ & 9.3$^{+0.07}_{-0.09}$ & 0.68$\pm$0.11 & 0.68$\pm$0.15\\ 
J0923+0402$^{\rm \ddagger}$ & 6.633$^{\rm \dagger}$ & 5940$\pm$327 & 3793$\pm$799 & 2682$\pm$135 & 46.5$^{+0.06}_{-0.07}$ & 46.79$^{+0.07}_{-0.08}$ & 47.51$^{+0.07}_{-0.08}$ & 9.11$^{+0.07}_{-0.08}$ & 9.42$^{+0.16}_{-0.24}$ & 1.92$\pm$0.45 & 0.95$\pm$0.44\\ 
PSOJ158-14 & 6.0685$^{\rm \dagger}$ & 6323$\pm$1538 & 3258$\pm$212 & 1767$\pm$1138 & 46.77$^{+0.06}_{-0.06}$ & 46.85$^{+0.1}_{-0.13}$ & 47.56$^{+0.1}_{-0.13}$ & 9.52$^{+0.38}_{-0.38}$ & 9.31$^{+0.07}_{-0.09}$ & 0.85$\pm$1.19 & 1.37$\pm$0.42\\ 
PSOJ183+05 & 6.4386$^{\rm \dagger}$ & 7075$\pm$2010 & 4476$\pm$1282 & 3035$\pm$348 & 46.69$^{+0.05}_{-0.06}$ & 46.49$^{+0.16}_{-0.25}$ & 47.2$^{+0.16}_{-0.25}$ & 9.29$^{+0.21}_{-0.42}$ & 9.41$^{+0.21}_{-0.41}$ & 0.62$\pm$0.47 & 0.48$\pm$0.36\\ 
PSOJ183-12$^{\rm \ddagger a)}$ & 5.86 & 6196$\pm$517 & 3203$\pm$517 & 3031$\pm$194 & 46.81$^{+0.02}_{-0.02}$ & 46.7$^{+0.06}_{-0.07}$ & 47.41$^{+0.06}_{-0.07}$ & 9.24$^{+0.08}_{-0.1}$ & 9.22$^{+0.12}_{-0.17}$ & 1.13$\pm$0.29 & 1.19$\pm$0.43\\ 
PSOJ217-16 & 6.1498$^{\rm \dagger}$ & 10292$\pm$909 & 2772$\pm$741 & 3243$\pm$1394 & 46.6$^{+0.04}_{-0.04}$ & 46.55$^{+0.08}_{-0.1}$ & 47.26$^{+0.08}_{-0.1}$ & 9.54$^{+0.27}_{-0.92}$ & 9.02$^{+0.19}_{-0.34}$ & 0.41$\pm$0.37 & 1.34$\pm$0.79\\ 
PSOJ217-07$^{\rm \ddagger}$ & 6.1663 & 9174$\pm$2402 & 2607$\pm$533 & 3260$\pm$2169 & 46.46$^{+0.07}_{-0.08}$ & 46.42$^{+0.15}_{-0.23}$ & 47.13$^{+0.15}_{-0.23}$ & 9.35$^{+0.39}_{-0.39}$ & 8.9$^{+0.16}_{-0.27}$ & 0.46$\pm$0.69 & 1.3$\pm$0.8\\ 
PSOJ231-20$^{\rm \ddagger}$ & 6.5869$^{\rm \dagger}$ & 6470$\pm$241 & 4644$\pm$179 & 2528$\pm$116 & 46.74$^{+0.01}_{-0.01}$ & 46.65$^{+0.04}_{-0.04}$ & 47.36$^{+0.04}_{-0.04}$ & 9.34$^{+0.05}_{-0.06}$ & 9.52$^{+0.04}_{-0.04}$ & 0.79$\pm$0.12 & 0.53$\pm$0.07\\ 
J1535+1943 & 6.370$^{\rm \dagger}$ & 6268$\pm$654 & 5640$\pm$236 & 2353$\pm$271 & 46.51$^{+0.11}_{-0.15}$ & 46.88$^{+0.1}_{-0.13}$ & 47.6$^{+0.1}_{-0.13}$ & 9.23$^{+0.13}_{-0.18}$ & 9.8$^{+0.06}_{-0.07}$ & 1.78$\pm$0.77 & 0.48$\pm$0.15\\ 
PSOJ239-07$^{\rm \ddagger}$ & 6.1102$^{\rm \dagger}$ & 4863$\pm$185 & 3947$\pm$79 & 537$\pm$63 & 46.72$^{+0.05}_{-0.06}$ & 46.74$^{+0.1}_{-0.13}$ & 47.46$^{+0.1}_{-0.13}$ & 9.68$^{+0.1}_{-0.13}$ & 9.42$^{+0.05}_{-0.06}$ & 0.46$\pm$0.17 & 0.83$\pm$0.24\\ 
PSOJ242-12$^{\rm a)}$ & 5.830 & 6791$\pm$1672 & 4892$\pm$495 & 1152$\pm$439 & 46.48$^{+0.07}_{-0.08}$ & 46.55$^{+0.12}_{-0.17}$ & 47.26$^{+0.12}_{-0.17}$ & 9.61$^{+0.28}_{-1.05}$ & 9.51$^{+0.1}_{-0.13}$ & 0.34$\pm$0.33 & 0.43$\pm$0.18\\ 
PSOJ308-27 & 5.7985 & 5284$\pm$148 & 2852$\pm$131 & 535$\pm$115 & 46.7$^{+0.02}_{-0.02}$ & 46.64$^{+0.06}_{-0.07}$ & 47.35$^{+0.06}_{-0.07}$ & 9.74$^{+0.16}_{-0.25}$ & 9.09$^{+0.05}_{-0.05}$ & 0.32$\pm$0.14 & 1.4$\pm$0.26\\ 
PSOJ323+12 & 6.5872$^{\rm \dagger}$ & 2828$\pm$140 & 2450$\pm$284 & 326$\pm$27 & 46.65$^{+0.02}_{-0.02}$ & 46.56$^{+0.06}_{-0.07}$ & 47.27$^{+0.06}_{-0.07}$ & 9.26$^{+0.08}_{-0.09}$ & 8.92$^{+0.09}_{-0.12}$ & 0.78$\pm$0.19 & 1.73$\pm$0.5\\ 
VIK J2211-3206$^{\rm \ddagger}$ & 6.3394$^{\rm \dagger}$ & 5114$\pm$224 & 3448$\pm$729 & 1811$\pm$97 & 46.66$^{+0.02}_{-0.02}$ & 46.73$^{+0.04}_{-0.05}$ & 47.44$^{+0.04}_{-0.05}$ & 9.27$^{+0.06}_{-0.07}$ & 9.3$^{+0.15}_{-0.24}$ & 1.15$\pm$0.2 & 1.06$\pm$0.47\\ 
VDES J2250-5051$^{\rm \ddagger}$ & 5.9767 & 16618$\pm$11883 & 5212$\pm$4022 & 3001$\pm$9685 & 46.58$^{+0.04}_{-0.05}$ & 46.73$^{+0.07}_{-0.08}$ & 47.44$^{+0.07}_{-0.08}$ & 9.99$^{+0.88}_{-0.88}$ & 9.66$^{+0.41}_{-0.41}$ & 0.22$\pm$1.45 & 0.46$\pm$0.72\\ 
SDSSJ2310+18$^{\rm \ddagger}$ & 6.0031$^{\rm \dagger}$ & 8576$\pm$2362 & 5156$\pm$252 & 3224$\pm$1504 & 46.92$^{+0.03}_{-0.04}$ & 46.78$^{+0.1}_{-0.13}$ & 47.49$^{+0.1}_{-0.13}$ & 9.55$^{+0.32}_{-0.32}$ & 9.67$^{+0.06}_{-0.08}$ & 0.67$\pm$0.75 & 0.51$\pm$0.15\\ 
PSOJ359-06 & 6.1722$^{\rm \dagger}$ & 3257$\pm$142 & 2653$\pm$213 & 554$\pm$39 & 46.55$^{+0.07}_{-0.08}$ & 46.59$^{+0.13}_{-0.19}$ & 47.3$^{+0.13}_{-0.19}$ & 9.23$^{+0.07}_{-0.09}$ & 9.0$^{+0.09}_{-0.12}$ & 0.9$\pm$0.36 & 1.53$\pm$0.65\\ 
\hline 
SDSSJ0100+28 & 6.3268$^{\rm \dagger}$ & 6647$\pm$1971 & 5742$\pm$1705 & 2496$\pm$316 & 47.58$^{+0.01}_{-0.01}$ & 47.44$^{+0.04}_{-0.04}$ & 48.15$^{+0.04}_{-0.04}$ & 9.82$^{+0.22}_{-0.45}$ & 10.1$^{+0.2}_{-0.1}$ & 2.94$\pm$1.48 & 0.77$\pm$0.52\\ 
ATLASJ025-33 & 6.3373$^{\rm \dagger}$ & 6408$\pm$962 & 3302$\pm$768 & 2461$\pm$251 & 47.03$^{+0.01}_{-0.01}$ & 46.95$^{+0.02}_{-0.02}$ & 47.66$^{+0.02}_{-0.02}$ & 9.51$^{+0.13}_{-0.2}$ & 9.37$^{+0.17}_{-0.1}$ & 0.88$\pm$0.73 & 0.21$\pm$0.7\\ 
ULASJ0148+06 & 5.977 & 5811$\pm$368 & 4741$\pm$473 & 2906$\pm$129 & 46.85$^{+0.01}_{-0.01}$ & 46.74$^{+0.04}_{-0.04}$ & 47.46$^{+0.04}_{-0.04}$ & 9.23$^{+0.06}_{-0.07}$ & 9.58$^{+0.08}_{-0.06}$ & 0.7$\pm$0.26 & 0.68$\pm$0.13\\ 
PSOJ036+03 & 6.5405$^{\rm \dagger}$ & 10131$\pm$329 & 3872$\pm$367 & 3727$\pm$135 & 46.82$^{+0.02}_{-0.02}$ & 46.79$^{+0.04}_{-0.05}$ & 47.5$^{+0.04}_{-0.05}$ & 9.55$^{+0.04}_{-0.05}$ & 9.43$^{+0.08}_{-0.07}$ & 0.51$\pm$0.32 & 0.66$\pm$0.2\\ 
QSOJ0439+1634$^{\rm c)}$ & 6.5188$^{\rm \dagger}$ & 5352$\pm$110 & 3329$\pm$295 & 1773$\pm$58 & 47.51$^{+0.02}_{-0.02}$ & 47.62$^{+0.03}_{-0.03}$ & 48.33$^{+0.03}_{-0.03}$ & 9.77$^{+0.03}_{-0.04}$ & 9.72$^{+0.07}_{-0.09}$ & 0.63$\pm$0.3 & 1.1$\pm$0.62\\ 
SDSSJ0818+17 & 5.96 & 9869$\pm$805 & 5477$\pm$334 & 3727$\pm$329 & 46.86$^{+0.04}_{-0.04}$ & 46.85$^{+0.08}_{-0.1}$ & 47.56$^{+0.08}_{-0.1}$ & 9.55$^{+0.1}_{-0.12}$ & 9.76$^{+0.06}_{-0.07}$ & 0.4$\pm$0.26 & 0.6$\pm$0.13\\ 
SDSSJ0836+00 & 5.773 & 6908$\pm$196 & 3793$\pm$691 & 573$\pm$77 & 47.06$^{+0.0}_{-0.0}$ & 47.14$^{+0.0}_{-0.0}$ & 47.85$^{+0.0}_{-0.0}$ & 10.15$^{+0.11}_{-0.14}$ & 9.59$^{+0.13}_{-0.08}$ & 1.36$\pm$0.23 & 1.72$\pm$0.51\\ 
SDSSJ0927+20 & 5.7722$^{\rm \dagger}$ & 5480$\pm$732 & 3405$\pm$243 & 1785$\pm$166 & 46.55$^{+0.05}_{-0.06}$ & 46.37$^{+0.15}_{-0.24}$ & 47.08$^{+0.15}_{-0.24}$ & 9.28$^{+0.12}_{-0.18}$ & 9.11$^{+0.1}_{-0.09}$ & 0.48$\pm$0.52 & 1.01$\pm$0.36\\ 
SDSSJ1030+05 & 6.304 & 5002$\pm$391 & 3578$\pm$336 & 1092$\pm$92 & 46.64$^{+0.05}_{-0.05}$ & 46.61$^{+0.11}_{-0.14}$ & 47.32$^{+0.11}_{-0.14}$ & 9.45$^{+0.09}_{-0.12}$ & 9.27$^{+0.09}_{-0.09}$ & 0.55$\pm$0.15 & 0.57$\pm$0.31\\ 
SDSSJ1306+03 & 6.033$^{\rm \dagger}$ & 4567$\pm$688 & 3825$\pm$449 & 769$\pm$189 & 46.66$^{+0.02}_{-0.02}$ & 46.53$^{+0.05}_{-0.06}$ & 47.24$^{+0.05}_{-0.06}$ & 9.5$^{+0.2}_{-0.37}$ & 9.29$^{+0.09}_{-0.18}$ & 0.67$\pm$0.66 & 0.48$\pm$0.19\\ 
ULASJ1319+09 & 6.1347$^{\rm \dagger}$ & 6964$\pm$2636 & 4905$\pm$162 & 3150$\pm$2607 & 46.73$^{+0.02}_{-0.03}$ & 46.58$^{+0.07}_{-0.08}$ & 47.3$^{+0.07}_{-0.08}$ & 9.28$^{+0.45}_{-0.45}$ & 9.53$^{+0.05}_{-0.11}$ & 0.71$\pm$0.49 & 0.13$\pm$0.09\\ 
CFHQSJ1509-1 & 6.1225$^{\rm \dagger}$ & 5095$\pm$828 & 3586$\pm$707 & 1021$\pm$216 & 46.59$^{+0.03}_{-0.04}$ & 46.65$^{+0.06}_{-0.08}$ & 47.37$^{+0.06}_{-0.08}$ & 9.47$^{+0.19}_{-0.33}$ & 9.3$^{+0.15}_{-0.17}$ & 6.63$\pm$5.37 & 0.31$\pm$0.39\\ 
\hline       
\\[0.01mm]                 
\end{tabular}
\tablefoot{ $\dagger$ Redshift from [CII] emission line (see \citealt{dodorico2023}). 
$\ddagger$ Classified as a Broad Absorption Line (BAL) quasar \citep{bischetti2022}. \\
a) Note that the \mgii\,emission line in this spectrum is very close to/within a strong telluric feature, therefore it needs to be taken with caution.\\
b) Note that the \civ\,emission line region in this spectrum has low SNR, therefore the derived black holes and Eddington ratios need to be taken with caution.\\
c) This quasar is gravitationally lensed (\citealt{fan2019}): the values presented here are not corrected for magnification. This quasar was not included in the comparison with lower$-z$ sources in the discussion session.}
\end{sidewaystable*}
%-----------------------  

%-------------------------------------   
\begin{table*}
\caption{Mean, median and standard deviation of bolometric luminosities, black hole masses and Eddington ratios distribution shown in Figure \ref{fig:MBHLbolHisto}, for the low$-$ and high$-z$ quasars samples. We also report the p-values obtained with a Kolmogorov-Smirnov test.}             
\label{table:mean}      
\centering          
\begin{tabular}{l c c c c c c c}     % 3 columns 
\hline\hline       
                      % To combine 4 columns into a single one 
 & \multicolumn{3}{c}{High$-z$ quasars} & \multicolumn{3}{c}{Low$-z$ quasars} & {KS-test}\\
 & {mean} & {median} & {st.dev} & {mean} & {median} & {st.dev} & {p-value} \\
\hline   
{All Luminosity}\\
L$_{\rm bol}$ [$10^{47}$\,erg s$^{-1}$]   & 1.49  & 1.32  & 0.38  & 1.49  & 1.32 & 0.38  & 1  \\
M$_{\rm BH}$  [$10^{9}\, M_{\odot}$] & 1.79 & 1.40 & 0.45 & 2.44 & 1.80 & 0.35 & 2$\times$10$^{-4}$\\
$\lambda_{\mathrm{BH}}$              & 0.89  & 0.76  & 0.30  & 0.62 & 0.50 & 0.34  & 2$\times$10$^{-8}$ \\
\hline
{High luminosity: L$_{\rm bol} > 10^{47.17}$ erg s$^{-1}$ }\\
L$_{\rm bol}$ [$10^{47}$\,erg s$^{-1}$]  & 2.47  & 2.18  & 0.14  & 2.47  & 2.17 & 0.14  & 1.0 \\
M$_{\rm BH}$  [$10^{9}\, M_{\odot}$] & 2.74 & 2.18 & 0.25 & 3.19 & 2.53 & 0.28 & 0.47\\
$\lambda_{\mathrm{BH}}$              & 0.88  & 0.73  & 0.24 & 0.85 & 0.69 & 0.28 & 0.32 \\
\hline
{Medium luminosity: $10^{46.92} <$ L$_{\rm bol} < 10^{47.17}$ erg s$^{-1}$} \\
L$_{\rm bol}$ [$10^{47}$\,erg s$^{-1}$]  & 1.20  & 1.26  & 0.07  & 1.20 & 1.26 & 0.07   & 1.0 \\
M$_{\rm BH}$  [$10^{9}\, M_{\odot}$] & 1.61 & 1.28 & 0.27 & 2.38 & 1.81 & 0.30 &  1$\times$10$^{-4}$\\
$\lambda_{\mathrm{BH}}$              &  0.84 & 0.76 & 0.28 & 0.61 & 0.51 & 0.29  & 6$\times$10$^{-5}$ \\
\hline
{Low luminosity: L$_{\rm bol} < 10^{46.92}$ erg s$^{-1}$} \\
L$_{\rm bol}$ [$10^{47}$\,erg s$^{-1}$]     & 0.49  & 0.58  & 0.30  & 0.49  & 0.58 & 0.30 & 1.0  \\
M$_{\rm BH}$  [$10^{9}\, M_{\odot}$] & 0.52 & 0.52 & 0.45 & 1.46 & 1.03 & 0.35 & 10$^{-8}$ \\
$\lambda_{\mathrm{BH}}$              & 0.79  & 0.77  & 0.38  & 0.41 & 0.32 & 0.34  & 3$\times$10$^{-9}$ \\
\hline
\end{tabular}
\end{table*}
%----------

\begin{acknowledgements}
      % George Becker
      G.B. was supported by the National Science Foundation through grant AST-1751404.
      
      % Emanuele Farina
      E.P.F. is supported by the international Gemini Observatory, a program of NSF’s NOIRLab, which is managed by the Association of Universities for Research in Astronomy (AURA) under a cooperative agreement with the National Science Foundation, on behalf of the Gemini partnership of Argentina, Brazil, Canada, Chile, the Republic of Korea, and the United States of America.
      
      % Romain Meyer
      R.A.M. acknowledge support from the ERC Advanced Grant 740246 (Cosmic\_Gas). 

      % Stefano Carniani
      S.C. is supported by European Union’s HE ERC Starting Grant No. 101040227 - WINGS. 

      \noindent
      % ESO
      This work is based on observations collected at the European Southern Observatory under ESO large program 1103.A-0817(A). 
\end{acknowledgements}

% WARNING
%-------------------------------------------------------------------
% Please note that we have included the references to the file aa.dem in
% order to compile it, but we ask you to:
%
% - use BibTeX with the regular commands:
%   \bibliographystyle{aa} % style aa.bst
%   \bibliography{Yourfile} % your references Yourfile.bib
%
% - join the .bib files when you upload your source files
%-------------------------------------------------------------------

\bibliography{bib_aa}
\bibliographystyle{aa}

\end{document}